# Cryothermal Measurements of Variable-Emittance Coatings with Lower Phase Transition Temperatures for Space Thermal Control

Chloe Stoops,[#] Vishwa Krishna Rajan,[#] and Liping Wang*

School for Engineering of Matter, Energy and Transport, Arizona State University, Tempe, Arizona 85287, USA

[#] *Equal contributions*

** Corresponding author: liping.wang@asu.edu*

**Abstract**

Space thermal control is critically important to ensure proper operation of on-board equipment in a regulated temperature range. Passive thermal control with variable-emittance coatings (VECs) could help save power consumption in a dynamically changing space thermal environment. Vanadium dioxide ($VO_2$) based VECs have been studied for space thermal control but its intrinsic phase transition around 68°C limits its wider space applications where lower temperature ranges are expected. In this work, we experimentally demonstrate enhanced radiative heat dissipation in space-like thermal environment via cryothermal measurements with VECs of undoped and tungsten doped $VO_2$. The fabricated undoped VEC exhibits a large emittance change of 0.6 across the phase transition, while the 1 at.% tungsten doped one shows an appreciable emittance variable of 0.4 with phase transition temperature lowered by 25°C. A vacuum cryothermal setup is developed with a liquid nitrogen cooled coldfinger to mimic cold space thermal background and a custom-designed sample mount suspended by nylon wires. After careful calibration and validation, greatly enhanced radiative heat dissipation upon $VO_2$ phase transition up to 3.5 times with transition temperature lowered by 25°C from 1 at.% tungsten doping is clearly observed from the cryothermal tests. In the actual space thermal environment, radiative heat flux could further increase across phase transition from 160 $W/m^2$ to 650 $W/m^2$ with undoped $VO_2$ coating from 55°C to 80°C, and from 175 $W/m^2$ to 493 $W/m^2$ with 1 at.% tungsten doped VEC from 30°C to 55°C.

**Keywords:** space thermal control, vacuum thermal test, VEC, vanadium dioxide, thin-film doping.

## 1. Introduction

Spacecrafts undergo dramatic temperature changes from the harsh heat of the sun and the freezing cold of the shadows, and they must maintain proper operating temperatures throughout mission life. Therefore, thermal control is crucial to any spacecraft mission [1]. There are many methods for maintaining operational temperatures on a spacecraft, including but not limited to cryogenic coolers, thermal blankets, foam insulation, on-board heaters and radiative coatings [2,3]. Considering the changing thermal environment in space, variable-emittance coatings (VECs) [4] could be a viable solution for adaptive space thermal control to alleviate power consumption. When the heat load is low, VECs have low emittance to retain the heat for preventing the temperature below the desired range. When the heat load is high, these VECs could dissipate more heat with high emittance to cold space to avoid the temperature rising beyond the operational values. Several works have theoretically evaluated the adaptive space thermal control performance with VECs [5–7].

Passive VECs are usually made of thermochromic materials whose optical properties change with temperature without any external input, and vanadium dioxide ($VO_2$) attracts most attention due to its unique phase transition behavior [8,9]. Undoped $VO_2$ could experience insulator-to-metal phase transition around 68°C, which is higher than the temperature ranges involved in most space missions. $VO_2$ phase transition temperatures can be lowered via doping with metallic elements such as tungsten as one common method at a rate of -20~25°C per at.% [10–12], via atomic layer deposition [11], reactive sputtering [13], pulsed laser deposition [14,15], vapor transport [16,17], and well-controlled furnace oxidation/annealing process [18,19]. Nanophotonic structures have been used to develop $VO_2$-based VECs with and without doping, including micro/nanostructure patterned $VO_2$ coatings [15,20–24], nanoparticle dispersed composites [25], and popular planar metafilm coatings [26–34] in typical $VO_2$-dielectric-metal structure. Such a thin film stack could achieve high emittance at elevated temperatures with metallic $VO_2$ phase, only which excites Fabry-Perot (FP) cavity resonant absorption, to promote heat dissipation. Large emittance change from these planar VECs with undoped and doped $VO_2$ has been experimentally demonstrated with different resonant spacer material mostly as inorganic dielectrics [29–34] and recently as organic polymer [35] for cost-effective fabrication.

While quite a few research groups have experimentally tested the thermal performance of $VO_2$ or tungsten-doped $VO_2$ based VECs for self-adaptive thermal regulation in the outdoor ambient terrestrial settings [15,25,35–38], there are very few studies which have experimentally demonstrated the dynamic thermal control with $VO_2$ VECs in vacuum thermal environment [15,33,39–42]. Taylor et al. [33] demonstrated the dynamic heat dissipation of $VO_2$ FP-based VEC from a thermal vacuum test with room temperature thermal background. Morsy et al. [39] experimentally showed the thermal regulation of $VO_2$ film on Si wafer with Au back coating in cold thermal background cooled by icy water. Tang et al. [15] conducted vacuum thermal test for doped $VO_2$ metasurfaces with dry-ice cooled chamber wall at 195 K, which is much higher than the 3 K of cold space. In a vacuum cryostat with liquid-nitrogen cooled coldfinger at 80 K to properly resemble the cold space thermal environment, Taylor et al. [40] investigated the dynamic heat dissipation and Boman et al. [41] studied the transient behaviors of the $VO_2$ FP-based VEC in a space-like environment. By using the same setup with acrylic plate supported sample mount, Wang et al. [42] experimentally demonstrated reduced temperature swing and rectified radiative heat transfer with the undoped $VO_2$ VEC for space thermal control.

In this work, we present the experimental observations of enhanced radiative heat dissipation with lower phase transition temperatures from VECs made of undoped and tungsten doped $VO_2$ from vacuum cryothermal tests. $VO_2$ based VECs are carefully designed with micrometer-thick polymers for the Fabry-Perot cavity spacer and the anti-reflection layer. Samples of 1 $\text{inch}^2$ are fabricated from thin film deposition, spin coating, and low-oxygen furnace process, while the temperature-dependent optical properties are characterized with infrared spectrometry. A custom-designed sample mount suspended by nylon wires, aiming to reach steady state faster than previous one supported by copper or acrylic plate, is used with integration of heat flux sensor, temperature sensor and thin-film heater. Careful calibrations for the sensitivity and parasitic heat loss of the heat flux sensor are carried out with black Actar and tungsten mirror samples, while validation measurements are performed with multiple static-emittance samples. Fabricated VEC samples made of undoped and 1 at.% tungsten doped $VO_2$ are measured with the cryothermal tests at multiple heating powers, to demonstrate the enhanced radiative heat dissipation and lowered transition temperatures by tungsten doping.

## 2. Coating Design and Optical Modeling

Figure 1(a) shows the design of proposed VECs comprising 50-nm $VO_2$ thin film, 1-μm polymer layer as FP cavity spacer and a 200-nm opaque aluminum layer deposited on the backside of 280-μm undoped silicon (UDSi) wafer. $VO_2$ can be doped with different amounts of tungsten to lower its phase transition temperature, while an anti-reflection (AR) layer can be added on top side of UDSi to possibly achieve larger emittance change. Spectral normal emittance is calculated for the proposed structure when the undoped $VO_2$ is in the insulating and metallic phases, as shown in Fig. 1(b). Without the AR coating, the spectral emittance is high due to strong FP resonance inside the polymer spacer with metallic $VO_2$ but is limited to about 0.7 due to the large reflection at the bare UDSi top surface. With the addition of the 1.5-μm polymer AR coating, spectral normal emittance is significantly increased up to 0.95 in the broadband that matches well with the spectral emissive power of a blackbody at 50°C. On the other hand, the AR coating little affects the low emittance spectrum of the proposed structure when the $VO_2$ is in insulating phase, which disables the strong resonance within the polymer cavity. Therefore, a larger emittance change could be achieved with the AR coating upon $VO_2$ phase transition. Moreover, as shown in Fig. 1(c), the spectral emittance exhibits some noticeable angular dependence with metallic $VO_2$ at emission angles greater than 45° because of strong FP resonance, while the proposed coating is quasi-diffuse with insulating $VO_2$.

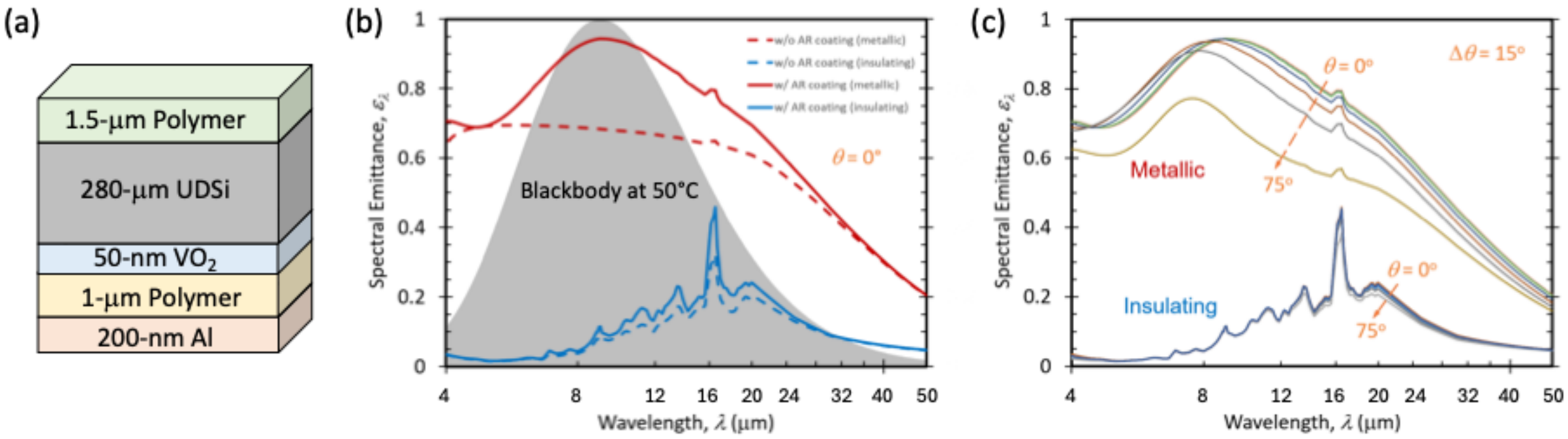

**Fig. 1.** (a) Design of variable-emittance coatings (VECs) made of 1.5-μm polymer as anti-reflection (AR) layer on 280-μm undoped silicon (UDSi) wafer whose backside is deposited with 50-nm undoped vanadium dioxide ($VO_2$), 1-μm polymer Fabry-Perot (FP) cavity spacer and 200-nm opaque Al films; (b) calculated spectral emittance ($\varepsilon_\lambda$) at normal direction ($\theta$=0°) for the VECs with and without the top polymer AR layer when $VO_2$ is in insulating and metallic phases; (c) calculated spectral emittance at multiple emission angles from 0° to 75° for the VECs with the AR layer when $VO_2$ is in insulating and metallic phases.

## 3. Coating Fabrication and Characterizations

The fabrication of the proposed variable emittance structures starts with sputtering 25-nm undoped vanadium or tungsten alloyed vanadium on to the double-side polished UDSi wafers with resistivity ρ > 10,000 Ω-cm, followed by thermal oxidation at 500 °C for 6 hours in a quartz tube furnace with $O_2$ content around 20 ppm with $N_2$ gas purging, while additional thermal annealing at 600°C is conducted for 1 at.% tungsten doped $VO_2$. Please refer to our previous work [19] about the details on the thermal growth of high-quality $VO_2$ and tungsten doped $VO_2$. AZ3312 photoresist as the polymer spacer is spin-coated on to the fully oxidized or annealed $VO_2$ layer with thickness to be 1 µm checked by a reflectrometer. The samples are made opaque by sputtering 200-nm aluminum onto the photoresist. Finally, Parylene-C is chosen as the AR coating and is deposited on the other bare side of UDSi wafers with about 1.5 µm thickness using SCS Parylene coater. The final variable-emittance samples are named "AR/UDSi/W0VO2FP" and "AR/UDSi/W1VO2FP" respectively for different tungsten doping.

The spectral reflectance ($R_\lambda$) for the fabricated structures is characterized by a Fourier-transform infrared spectrometer (FTIR) in the wavelength range from 4 to 24 µm equipped with a specular reflectance accessory at 10° angle of incidence. Samples are mounted on a home-built heating/cooling stage whose temperature is feedback controlled with ±1°C accuracy. Figures 2(a) and 2(b) show the temperature-dependent spectral reflectance measured during heating and cooling of AR/UDSi/W0VO2FP structure at 5°C intervals, respectively. Clearly, the spectral reflectance drops significantly when undoped $VO_2$ transits from insulator to metal. Note that the major reflectance dip around $\lambda$=16µm is due to the phonon absorption of UDSi wafer, while the minor dips elsewhere are mainly associated with the absorption by the polymers. The spectral emittance ($\varepsilon_\lambda=1-R_\lambda$) at wavelength $\lambda$=8µm as shown in Fig. 2(c) could increase from 0.2 to 0.9 upon phase transition between 60°C and 80°C, while the change is relatively smaller about 0.4 at other selected longer wavelengths of $\lambda$=16µm and 24µm. With 1 at.% tungsten doping, the AR/UDSi/W1VO2FP structure exhibits very similar behaviors with large variation in the temperature-dependent spectral reflectance upon heating and cooling respectively as shown in Figs. 2 (d) and 2(e), as well as the spectral emittance in Fig. 2(f) upon phase transition but at temperatures lower by about 20°C.

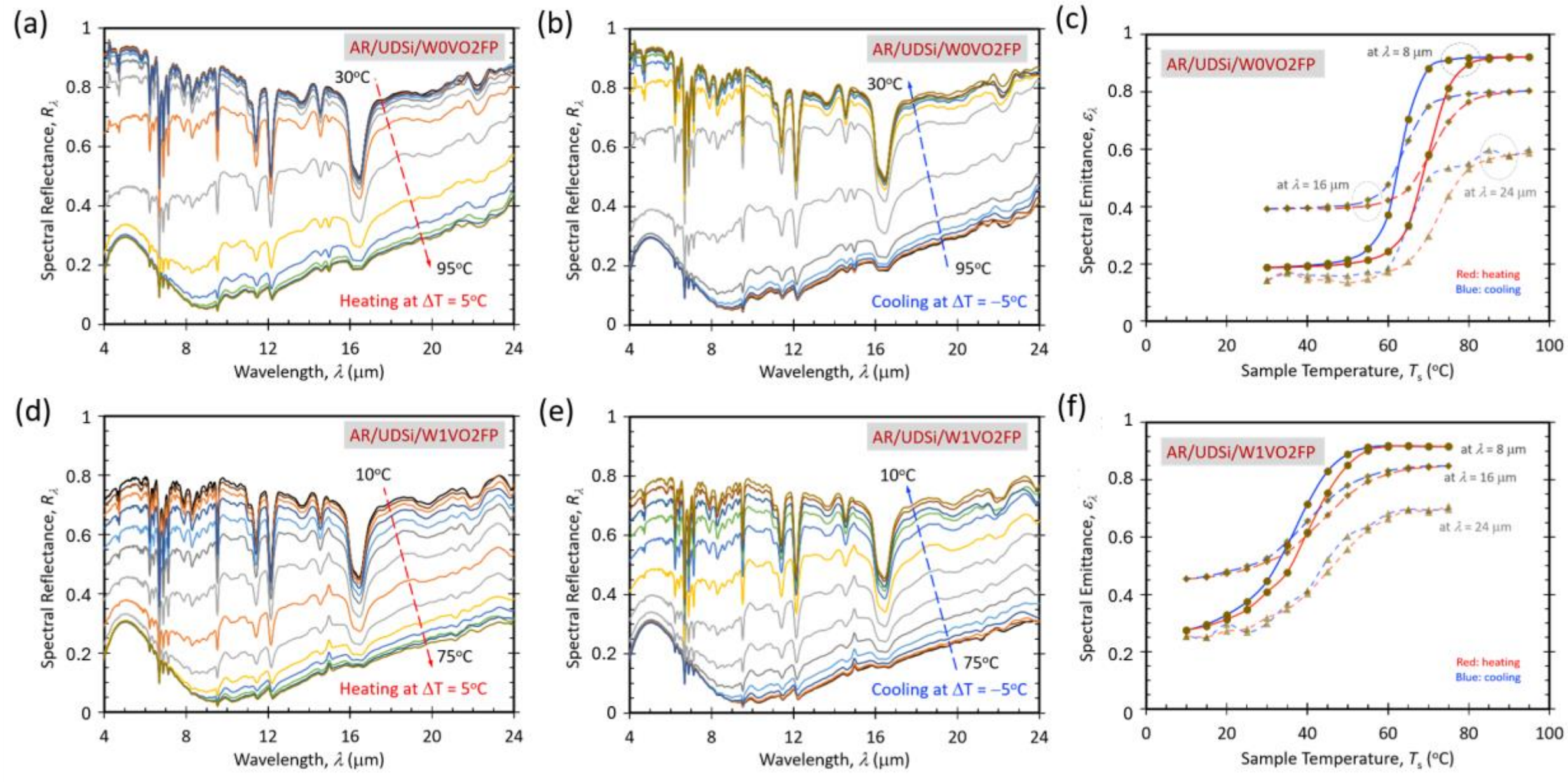

**Fig. 2.** Measured temperature-dependent spectral infrared reflectance at near-normal direction upon (a,d) heating and (b,e) cooling with temperature intervals of 5°C as well as (c,f) spectral emittance at selected infrared wavelengths for the fabricated variable-emittance coating samples with (a,b,c) undoped $VO_2$ (AR/UDSi/W0VO2FP) from 30°C to 95°C and (d,e,f) 1 at.% tungsten-doped $VO_2$ (AR/UDSi/W1VO2FP) from 10°C to 75°C.

Figures 3(a) and 3(b) present the total-normal emittance of the fabricated VECs with undoped and 1 at.% tungsten doped $VO_2$ as markers, which is calculated as $\varepsilon_{s,norm}(T_s) = \int_{4\mu m}^{24\mu m} \varepsilon_\lambda(T_s) E_{b,\lambda}(T_s) d\lambda / \int_{4\mu m}^{24\mu m} E_{b,\lambda}(T_s) d\lambda$ with spectral blackbody emissive power $E_{b,\lambda}$ is from Planck's law. The total-normal emittance of the AR/UDSi/W0VO2FP structure remains almost the same around 0.2 at temperatures below 55°C with insulating $VO_2$ phase. Upon phase transition, the total-normal emittance increases and reaches around 0.8 with metallic $VO_2$ around 80°C, leading to a large change of 0.6. On the other hand, with 1 at.% tungsten doping, the AR/UDSi/W1VO2FP structure exhibits similar large variation of total-normal emittance with 0.4 at 30°C and 0.8 at 55°C upon phase transition but at lowered temperatures by 25°C. In addition, the thermal hysteresis between heating and cooling processes is narrowed from ~12°C with undoped $VO_2$ to ~5°C with 1 at.% tungsten doping. To facilitate the heat transfer modeling for the cryothermal tests, the measured total-normal emittance of the VECs is fitted with four linear relations for the insulating phase, transition upon heating, transition upon cooling, and metallic phase, as labelled in the figures.

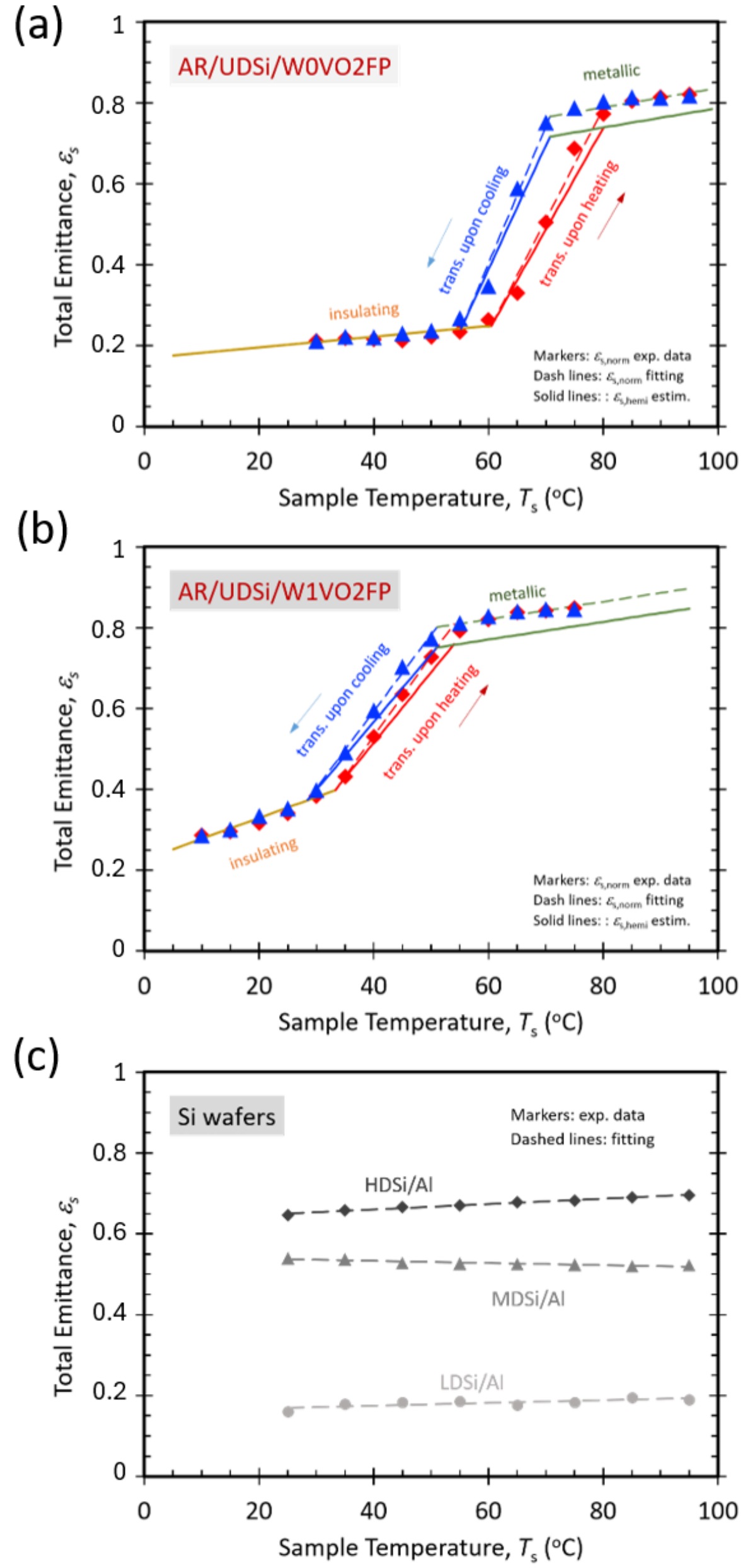

**Fig. 3.** Total emittance ($\varepsilon_s$) obtained from the measured temperature-dependent spectral emittance for the fabricated variable-emittance coatings (VECs) with (a) undoped $VO_2$ (AR/UDSi/W0VO2FP) and (b) 1 at.% tungsten-doped $VO_2$ (AR/UDSi/W1VO2FP), as well as (c) that for the heavily (HDSi/Al), medium (MDSi/Al) and lightly doped silicon (LDSi/Al) wafers with backside Al films as static-emittance samples. Experimental data for total normal emittance ($\varepsilon_{s,norm}$) are shown as markers with linear fittings as dashed lines. Based on the optical modeling for the VECs, the angular dependence could reduce the total hemispherical emittance ($\varepsilon_{s,hemi}$) by about 5% from the total normal emittance ($\varepsilon_{s,norm}$) when undoped and tungsten doped $VO_2$ are in metallic phase, while it has negligible effect at insulating phase. The total hemispherical emittance of the VECs is fitted in solid lines to be used for cryothermal modeling.

As the optical modeling suggested slight angular dependance only at metallic phase due to excitation of strong FP resonance, 5% reduction is estimated on the total-hemispherical emittance ($\varepsilon_{s,hemi}$ or simply $\varepsilon_s$) from the total-normal emittance ($\varepsilon_{s,norm}$), while both are considered the same at insulating phase. The linear relations are updated accordingly for the total-hemispherical emittance to be used in the theoretical modeling of cryothermal tests. For validating the cryothermal tests, a set of static-emittance samples made of lightly, medium, and heavily doped silicon wafers with backside coated with opaque Al film (i.e., HDSi/Al, MDSi/Al, LDSi/Al) are used. Figure 3(c) shows the total emittance of these silicon wafers from the room-temperature FTIR reflectance measurements, which exhibit little temperature dependence from 20°C to 100°C. Note that the thermal emission from Si wafers is reasonably quasi-diffuse.

## 4. Cryothermal Test Setup

In order to experimentally demonstrate the dynamic radiative heat dissipation with lowered phase transition temperature from the VEC made of tungsten doped $VO_2$ for space thermal control, cryothermal tests were conducted inside a cryostat (Janis VPF-800) under high vacuum ($< 1\times10^{-3}$ Pa) with a coldfinger maintained at 80 K by liquid nitrogen to mimic the space-like thermal environment. As shown in Fig. 4(a), the test sample, a heat flux sensor with $\pm$5% accuracy, and a polyimide thin-film heater, all in 1-inch squared size ($A_s$), are attached together with thermal paste. The stack is suspended by two thin nylon wires at a spacing from 7 to 9 mm to the coldfinger, aiming to reach steady state faster than our previous methods with copper or acrylic plate [44-46]. Black Actar with emissivity about 0.95 is attached onto the coldfinger and the backside of the thin-film heater, while a thermistor with an accuracy of $\pm$0.1°C is inserted between the sample and the heat flux sensor for measuring the sample temperature. Two digital multimeters measure the heat flux sensor voltage and thermistor resistance, while a programmable power supply controls the voltage supplied to the heater. The temperature of the coldfinger is monitored with an E-type thermocouple fixed at its backside. All the experimental data is recorded with a computer program during entire course of the cryothermal test. Figure 4(b) shows the view of the sample mounted inside the cryostat in high vacuum and filled with LN2, and an equivalent heat transfer model is depicted in Fig. 4(c), which is to be discussed later.

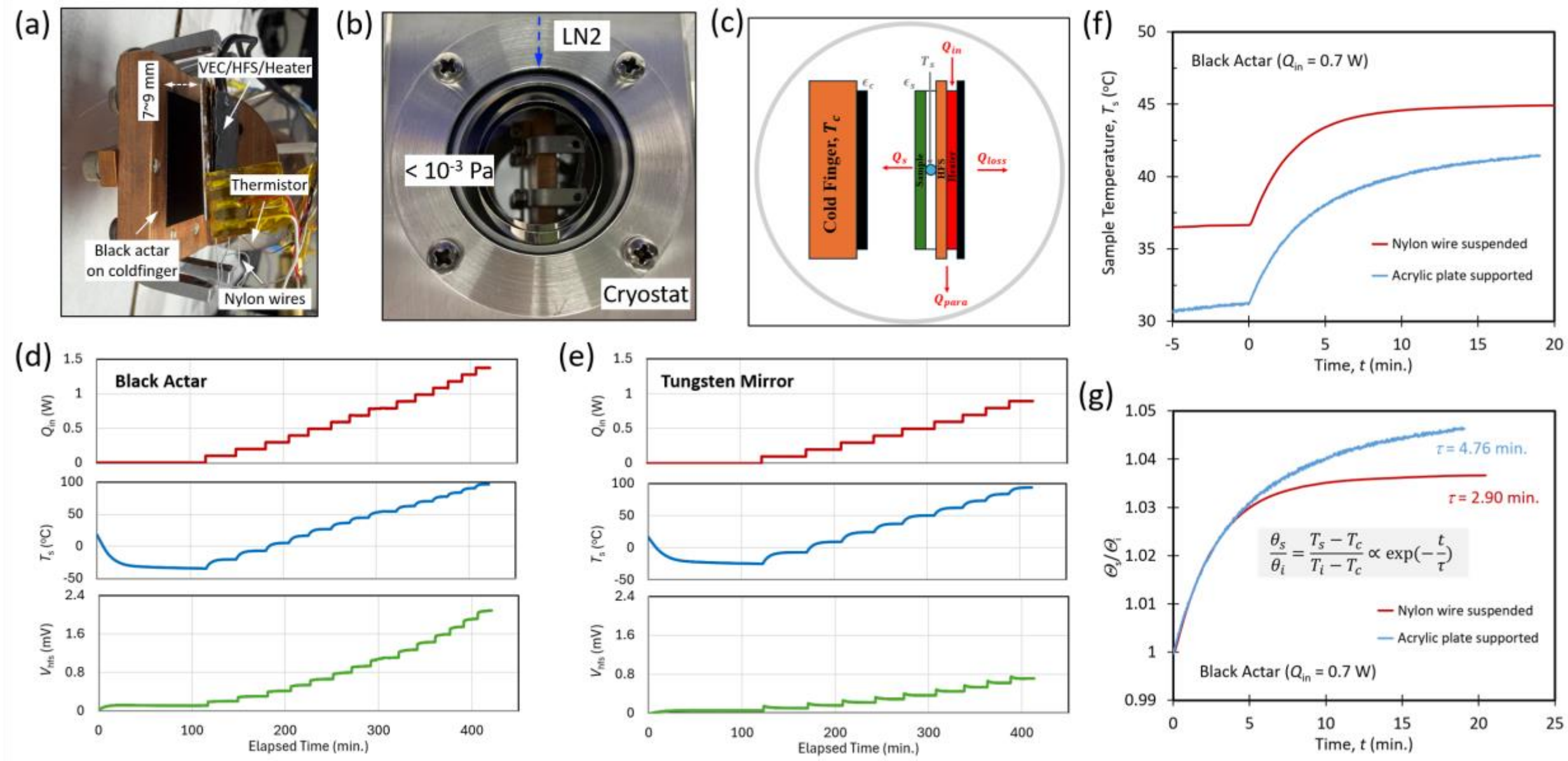

**Fig. 4.** (a) Photo of variable-emittance coating sample, heat flux sensor (HFS) and thin-film heater stacked with thermal paste and suspended by two Nylon wires above the coldfinger with a spacing of 7~9 mm, leading to a view factor of 0.52~0.60; (b) View of the sample mounted inside the cryostat pumped down ($P$ < 0.001 Pa) and filled with liquid nitrogen (LN2); (c) an equivalent heat transfer model for the cryothermal test; (d,e) step-wise heating process with heater power input, sample temperature and HFS voltage reading from the cryothermal tests (d) for black Actar ($\varepsilon_s$=0.95) and (e) for tungsten mirror ($\varepsilon_s$=0.03); (f) transient sample temperature ($T_s$) and (g) normalized temperature ratio ($\Theta_s/\Theta_i$) under the same heating input ($Q_{in}$=0.7 W) for black Actar suspended by nylon wires in this work and supported by acrylic plate in our previous work [46].

Figures 4(d) and 4(e) respectively present the cryothermal test processes for a black Actar sample ($\varepsilon_s$ = 0.95) and a highly reflective tungsten mirror ($\varepsilon_s$ = 0.03) with transient temperature and heat flux sensor voltage measured with step-wise heating power inputs. In particular, after the LN2 is filled, the cryostat is left for about 2 hours for the sample to reach the lowest steady-state temperature with heater power off. Then the heater is supplied with step-wise power input ($Q_{in}$), resembling net heating load of a spacecraft from both internal and external sources, at 0.1 W intervals until the sample finally reaches almost 100°C. At each heating power, sufficient time from 15 to 45 mins is given to ensure that the sample has reached steady state for at least 5 mins. In general, it would require more time to reach steady state at lower heating power for samples with smaller emissivity. Due to the large difference in emissivity, black Actar requires 1.4 W heating power to reach about 100°C, while tungsten mirror only needs 0.9 W.

To demonstrate if the current sample mounting method with nylon wires could reach steady state faster than previously used acrylic support plate, the measured transient temperature ($T_s$) of black Actar from both methods is compared in Fig. 4(f) at the same heating power of 0.7 W. To quantitatively find the thermal time constant $\tau$ from each method, normalized temperature difference, $\frac{\theta_s}{\theta_i} = \frac{T_s - T_c}{T_i - T_c}$, with initial temperature $T_i$ and coldfinger temperature $T_c$, is plotted in Fig. 4(g). Considering its linear relation to $\exp\left(-\frac{t}{\tau}\right)$ [43], the thermal time constant is fitted to be 2.90 mins for the nylon wire suspended sample mount, compared to 4.76 mins for the previous acrylic plate supported one, leading to about 40% faster in reaching steady state. In addition, the nylon wire suspended black sample achieved slightly higher temperature by about 3°C than the acrylic plate supported one with the same heating power of 0.7 W, indicating smaller heat loss ($Q_{\text{loss}}$) from the suspension method.

## 5. Calibration and Validation

To obtain the experimental radiative heat transfer ($Q_s$) by the sample to the coldfinger, representing the radiative heat dissipation to the cold outer space, the sensitivity ($S$) and parasitic heat loss ($Q_{\text{para}}$) from the heat flux sensor need to be calibrated first as a function of sample temperature ($T_s$). According to the heat transfer model in Fig. 4(c), energy balance yields the following at steady state:

$$Q_{in} = Q_s + Q_{para} + Q_{loss} \tag{1}$$

where the heater power input is found with voltage and current from the power supply as $Q_{in} = VI$. The experimental radiative heat transfer by the sample $Q_{s,exp}$ is:

$$Q_{s,exp} = \frac{V_{hfs}}{S(T_s)} - Q_{para}(T_s) \tag{2}$$

with the heat flux sensor voltage ($V_{\text{hfs}}$) and sample temperature ($T_s$) are taken as the average values from 300 data points during last 5-mins steady state at a given heater power input. On the other hand, the theoretical radiative heat transfer by the sample ($Q_{s,theo}$) to the coldfinger at temperature $T_c$ is given by

$$Q_{s,theo} = \epsilon_{\text{eff}} \sigma A_s (T_s^4 - T_c^4) \tag{3}$$

where $\varepsilon_{\text{eff}} = \left(\frac{1}{\varepsilon_s} + \frac{1}{F} + \frac{1}{\varepsilon_\text{c}} - 2\right)^{-1}$ is the effective emittance between two squared surfaces of same size, depending on sample total-hemispherical emittance $\varepsilon_s$, view factor $F$ between the sample and the coldfinger, and total emittance of the black-coated coldfinger ($\varepsilon_c$= 0.95). During the tests, the spacing between the sample and coldfinger is measured every time but it could vary from 7 to 9 mm among different sample mounting due to the flexible nature of nylon wires, leading to view factors from 0.52 to 0.60 [43]. As a result, the effective emittance is smaller than the sample emittance with non-unity view factors in particular for highly emissive samples.

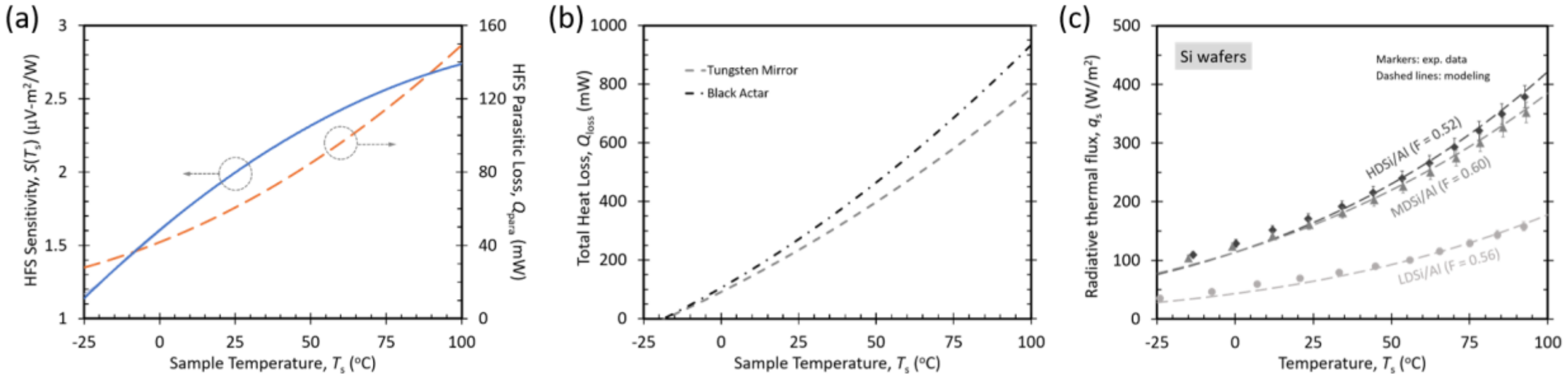

**Fig. 5.** (a) Calibrated sensitivity $S$ and parasitic heat loss $Q_{\text{para}}$ for the heat flux sensor (HFS) as a function of sample temperature based on the cryothermal tests for the black Actar and tungsten mirror samples. Second-order polynomial fitting is used for both to process the HFS voltage and temperature readings from the cryothermal tests to obtain the experimental radiative heat flux for other samples including the VECs and static emitters. (b) Total heat loss ($Q_{\text{loss}}$) from the tungsten mirror and black Actar as a function of sample temperature; (c) Cryothermal test validation with radiative heat flux ($q_s$) at different steady-state temperatures from several static-emittance samples: HDSi/Al (with view factor $F$ = 0.52), MDSi/Al ($F$=0.60), and LDSi/Al ($F$=0.56), where cryothermal measurements are shown as markers with 5% uncertainty along with theoretical modeling as dashed lines.

By setting the experimental and theoretical heat fluxes equal, i.e., $Q_{s,theo} = Q_{s,exp}$, for the black Actar and tungsten mirror, the two unknown parameters, sensitivity ($S$) and parasitic heat loss ($Q_{\text{para}}$) of the heat flux sensor, can be found with the steady state data from the cryothermal tests at each heater power input. Their dependences in a wide range of sample temperatures from –25°C to 100°C are shown in Fig. 5(a), and second order polynomial fitting is used for both to obtain analytical relations with R squared values greater than 0.999 for facilitating the heat transfer modeling. The total heat loss ($Q_{loss}$) from the front and back surfaces of the sample mount is calculated based on Eq. (1) for the black Actar and tungsten mirror samples, as shown in Fig. 5(b). The heat loss could be appreciable, up to 0.93 W at 100°C with the black sample, which is about 65% of the total heater input. This is understandable as the majority of the heat loss is from the back of the sample mount, which is covered by black Actar, to the surroundings at about room temperature. While the heat loss on the back side can be minimized by aluminum foil, but the time to reach steady state could increase significantly [44,45]. Tungsten mirror shows a relatively smaller heat loss, thanks to its low emissivity, which minimizes the heat loss to the surroundings on the sample side.

Finally, the cryothermal tests are validated with three static silicon wafers of different doping, whose emittance is characterized as shown in Fig. 3(c). Following the same test procedures, the experimental radiative heat flux, $q_s = Q_s/A_s$, is obtained from the measured steady-state heat flux sensor voltage ($V_{hfs}$) and sample temperature ($T_s$) along with calibrated sensitivity ($S$) and parasitic heat loss ($Q_{\text{para}}$) according to Eq. (2). As shown in Fig. 5(c), the processed experimental data with 5% uncertainty as markers agrees well with the modeling for all three silicon wafers of different emittance values, which undoubtedly verifies the cryothermal tests. In particular, HDSi/Al exhibits almost the same radiative heat flux from the cryothermal test as the MDSi/Al, though their emittance could differ by up to 0.2. This is because of the difference in view factor, which is $F$=0.60 with ~7 mm spacing for MDSi/Al compared to 0.52 with ~9 mm spacing for HDSi/Al, due to sample mounting with flexible Nylon wires. On the other hand, LDSi/Al has a view factor of 0.56 with ~8 mm spacing to the coldfinger during the test.

## 6. Cryothermal Tests with Variable-emittance Coatings

Fabricated VEC samples made of undoped and 1 at.% tungsten doped $VO_2$ are tested with the cryothermal setup. The radiative heat flux as a function of the sample temperature is presented in Fig. 6 with both measured ones as markers and theoretical validation in dashed lines for the cryothermal tests. For the undoped VEC sample (AR/UDSi/W0VO2FP) in Fig. 6(a), enhanced radiative heat dissipation is clearly observed across the phase transition by 3.5 times from 89 W/m$^2$ at 59°C to 313 W/m$^2$ at 80°C upon heating (or 113 W/m$^2$ at 57°C to 341 W/m$^2$ at 78°C upon cooling), with good agreement between measurements and modeling. Note that during the cryothermal test the undoped sample is mounted about 9 mm away from the coldfinger, leading to a view factor $F$ = 0.52. However, in the actual space environment, the VEC has full view with $F$ = 1 to cold 3 K thermal background, and the modeling predicts a larger radiative heat transfer enhancement by 4.1 times from 160 W/m$^2$ to 650 W/m$^2$ upon phase transition of undoped $VO_2$ from 55°C to 80°C.

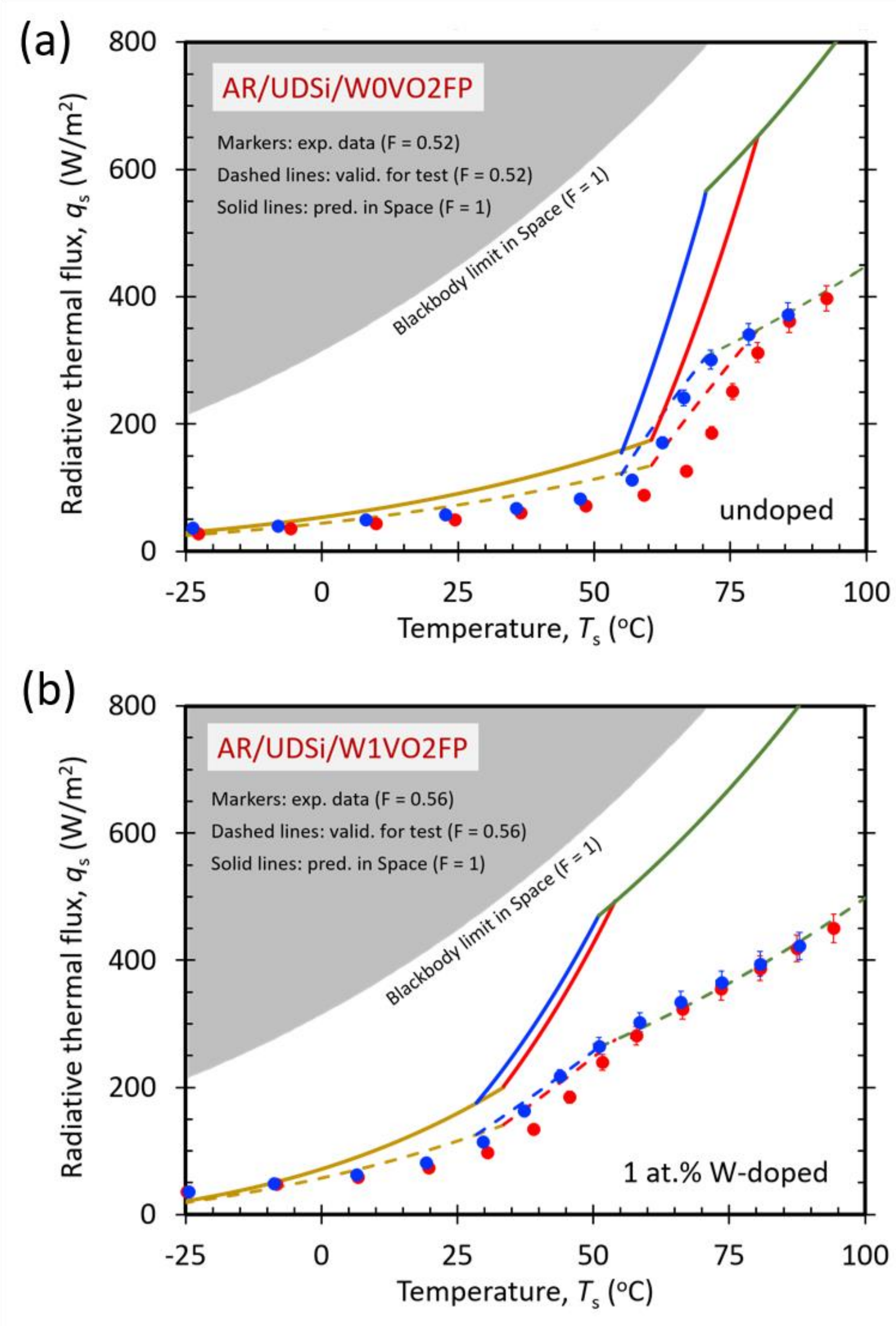

**Fig. 6.** Experimental demonstration of tunable radiative heat dissipation from the variable-emittance coatings (VECs): (a) with undoped $VO_2$ (at a view factor of $F$ = 0.52 to the black-coated coldfinger) and (b) 1 at.% tungsten-doped $VO_2$ (at $F$ = 0.56) with lower phase transition temperatures. Cryothermal measurements are shown as markers with error bars for 5% uncertainty. The modeling with the same view factors is depicted as dashed lines for validating the cryothermal tests, while that with unity view factor ($F$=1) are calculated as solid lines to predict the radiative heat dissipation in outer Space with full view to 3 K cold background for the VECs along with blackbody limit.

With 1 at.% tungsten doping, the variation emittance coating sample exhibits enhanced radiative heat dissipation at lowered temperatures by about 25°C, as shown in Fig. 6(b). In particular, the radiative heat flux measured from the cryothermal test with a view factor of $F$ = 0.56 demonstrates a 2.5-fold enhancement across the phase transition from 98 W/m$^2$ at 31°C to 241 W/m$^2$ at 52°C upon heating (or 114 W/m$^2$ at 30°C to 265 W/m$^2$ at 51°C upon cooling), which matches with modeling validation. The radiative heat dissipation is predicted to vary by 2.8 times from 175 W/m$^2$ to 493 W/m$^2$ upon phase transition from 30°C to 55°C with 1 at.% tungsten doped VEC in the space environment, while the heat flux could go up to 658 W/m$^2$ at 55°C if the coating could reach unity emittance as black surface.

## 7. Conclusion

In summary, $VO_2$ based variable-emittance coatings have been carefully designed with polymer cavity spacer and polymer anti-reflection coatings, and successfully fabricated with thin-film deposition and furnace oxidation methods. The undoped one achieved a large emittance change of 0.6 from 55°C to 80°C upon phase transition, while 1 at.% tungsten doping resulted in the phase transition temperature to be lowered by about 25°C and emittance variation of 0.4. Cryothermal setup with custom designed sample mount suspended by nylon wires has been developed with careful calibration and validation from static-emittance samples, with steady state reached 40% faster than previous acrylic plate supported sample mount. Cryothermal tests for the VEC samples evidently demonstrated greatly enhanced radiative heat dissipation upon $VO_2$ phase transition up to 3.5 times with the transition temperature lowered by 25°C with 1 at.% tungsten doping. It is predicted that the enhancement could reach 4.1 times with undoped coating and 2.8 times with 1 at.% tungsten doping in the actual space thermal environment across the phase transition. Future work would involve further optimization of the VECs to achieve emittance variation greater than 0.6, as well as more tungsten doping for further lowered phase transition, to be experimentally demonstrated with the cryothermal measurements in space-like vacuum thermal environment.

## Conflict of Interest

The authors have no conflicts to disclose.

## Author Contributions

C.S. carried out the cryothermal tests and conducted heat transfer modeling; V.K.R. designed, fabricated and characterized the samples; C.S. and L.W. analyzed the data; C.S. and V.K.R. wrote the initial manuscript draft; L.W. prepared the figures, revised the manuscript, conceived the idea, secured funds, and supervised the project; all authors reviewed the final manuscript.

## Acknowledgements

L.W. and V.K.R. would like to thank the support from National Science Foundation (CBET-2212342). C.S. is grateful for the support from the Master's Opportunity for Research in Engineering (MORE) program, the Fulton Undergraduate Research Initiative (FURI) program, and Barrett, The Honors College at Arizona State University.

## Data Availability

The data that support the findings of this study are available from the corresponding author upon reasonable request.